\documentclass[a4paper,12pt]{article}%
\usepackage[english]{babel}
\usepackage[ansinew]{inputenc}
\usepackage{enumerate}
\usepackage{lmodern}
\usepackage{setspace}  
\usepackage{epsfig}%
\usepackage{amsmath}%
\usepackage{amsfonts}%
\usepackage{dsfont}
\usepackage{mathrsfs}%
\usepackage{amssymb}%
\usepackage{graphicx}%
\usepackage{undertilde}%
\usepackage{longtable}%
\usepackage{geometry, calc, color, setspace}%
\usepackage{indentfirst}%
\usepackage{wrapfig}%
\usepackage{float}
\usepackage{scalefnt}
\usepackage{rotating}
\usepackage{lscape}
\usepackage{fullpage}
\usepackage[nottoc]{tocbibind}          
\usepackage{natbib} 
\newcommand{\bff}[1]{\boldsymbol{#1}}

\def\ds{\displaystyle}

\begin{document}

\title{{Bayesian linear regression models with flexible error distributions}}


\author{N\'ivea B. da Silva$^{\rm 1}$\thanks{Email address: nivea.bispo@gmail.com; moprates@gmail.com; fbgoncalves@est.ufmg.br.
    \vspace{6pt}}, Marcos O. Prates$^{\rm 1}$ and Flávio B. Gonçalves$^{\rm 1}$\\ \vspace{6pt}
    $^{\rm 1}${{\small Department of Statistics, Federal University of Minas Gerais, Brazil}}\\
}

\maketitle

\begin{abstract}
\noindent This work introduces a novel methodology based on finite mixtures of Student-t distributions to model the errors' distribution in linear regression models. The novelty lies on a particular hierarchical structure for the mixture distribution in which the first level models the number of modes, responsible to accommodate multimodality and skewness features, and the second level models tail behavior. Moreover, the latter is specified in a way that no degrees of freedom parameters are estimated and, therefore, the known statistical difficulties when dealing with those parameters is mitigated, and yet model flexibility is not compromised. Inference is performed via Markov chain Monte Carlo and simulation studies are conducted to evaluate the performance of the proposed methodology. The analysis of two real data sets are also presented.

\end{abstract}

\section{Introduction}\label{sec:1}

Density estimation and modeling of heterogeneous populations through finite mixture models have been highly explored in the literature \citep{sylvia2006finite,mclachlan2000finite}. In linear regression models, the assumption of identically distributed errors is not appropriate for applications where unknown heterogeneous latent groups are presented. A simple and natural extension to capture mean differences between the groups would be to add covariates in the linear predictor capable of characterizing them. However, that may not be enough to explain the source of heterogeneity often presented in the data. Furthermore, differences may also be presented in skewness, variance and tail behavior. Naturally, the usual normality assumption is not appropriate in those cases.

Over the years, many extensions of the classical normal linear regression model, such the Student-t regression \citep{lange1989}, have been proposed. In practice, the true distribution of the errors is unknown and it may be the case that single parametric family is unable to satisfactorily model their behavior. One possible solution is to consider a finite mixture of distributions, which are a natural way to detect and model some unobserved heterogeneity. Conceptually, this may be seen as a semiparametric approach for linear regression modelling. Initial works in this direction were proposed in \citet{bartolucci2005} and \citet{soffritti2011multivariate} by assuming a mixture of Gaussian distributions to model the errors. A more flexible approach was presented in \citet{galimberti2014multivariate} with a finite mixture of (multivariate) Student-t distributions.

From a modelling perspective, mixtures of more flexible distributions like the Student-t, skew-normal, skew-t, may be preferred over mixtures of Gaussians when the data exhibit mutimodality with significant departure from symmetry and Gaussian tails. Basically, the number of Gaussian components in the mixture to achieve a good fit may defy model parsimony. On the other hand, more flexible distributions usually have specific parameters with statistical properties that hinder the inference procedure \citep[see][]{1999fernandez, fonseca2008objective, villa2014objective}. A common solution for that is to impose restrictions on the parametric space, which goes in the opposite direction of model flexibility. For example, in \citet{galimberti2014multivariate}, inference is performed via maximum likelihood using the EM algorithm with restrictions on the degrees of freedom parameters.

The aim of this paper is to propose a flexible model for the errors in a linear regression context but that, at the same time, is parsimonious and does not suffer from the known inference problems related to some of the parameters in the distributions described above. Flexibility and parsimony are achieved by considering a finite mixtures of Student-t distributions that, unlike previous works, considers a separate structure to model multimodality/skewness and tail behavior. Inference problems are avoided by a model specification that does not require the estimation of the degrees of freedom parameter and, at the same time, does not lose model flexibility. That is achieved due to the fact that arbitrary tail behaviors of the Student-t can be well mimicked by mixtures of well-chosen Student-t distributions with fixed degrees of freedom. Inference for the proposed model is performed under the Bayesian paradigm through an efficient MCMC algorithm.

This paper is organised as follows: the proposed model is presented in Section \ref{sec:2} and the MCMC algorithm for inference is presented in Section \ref{sec:3} along with the discussion of some computational aspects of the algorithm. Simulated examples are presented in Section \ref{sec:4} and the results of the analysis of two real data sets are shown in Section \ref{sec:5}. Finally, Section \ref{sec:6} presents some concluding remarks and proposals of future works.

\section{Linear regression with flexible errors's distribution}\label{sec:2}

\subsection{Motivation}\label{sec:2.1}

Finite mixture models have great flexibility to capture specific properties of real data such as multimodality, skewness and heavy tails and has recently been used in a linear regression context to model the errors' distribution. \citet{bartolucci2005} and \citet{soffritti2011multivariate} were the first to consider a finite mixture to model the errors in linear regression models. More specifically, they proposed a finite mixture of $d$-dimensional Gaussian distributions. \citet{galimberti2014multivariate} extended that model by assuming that the errors $\epsilon_i$ follow a $d$-dimensional Student-t distribution, i.e.
\begin{eqnarray*}
f(\epsilon_{i}) = \displaystyle \sum_{j=1}^J w_jf_{\mathcal{T}_d}(\mu_j, \Sigma_j,\nu_j),
\end{eqnarray*}
where $f_{\mathcal{T}_d}$ denotes the density of the $d$-dimensional Student-t distribution with mean vector $\bff{\mu}$, positive definite dispersion matrix $\bff{\Sigma}$ and degrees of freedom $\bff{\nu}$.

Recently, \citet{benites2016} considered regression models in which the errors follow a finite mixture of scale mixtures of skew-normal (SMSN) distributions, i.e.
\begin{eqnarray*}
f(\epsilon_{i}) = \displaystyle \sum_{j=1}^J w_jf_{SMSN}(\mu_j, \sigma^2_j,\lambda_j,\nu_j).
\end{eqnarray*}
where $f_{SMSN}$ corresponds to some distribution that belongs to the SMSN class, with location parameter $\mu_j$, dispersion $\sigma^2_j$, skewness $\lambda_j$ and degree of freedom $\nu_j$, for $j=1,\ldots,J$.

Mixtures of Gaussian distributions may not be the most adequate choice when the error terms present heavy tails or skewness. The simultaneous occurrence of multimodality, skewness and heavy tails may require a large number of components in the mixture which, in turn, may compromise model parsimony. The mixture models proposed by \citet{galimberti2014multivariate} and \citet{benites2016} are more flexible and remedies the problem of dealing with outliers, heavy tails and skewness in the errors. However, both models have a constraint with respect to the estimation of the degrees of freedom parameters, once the estimation of these parameters are known to be difficult and computationally expensive. Thus, for computational convenience, they assume that $\nu_1=\ldots=\nu_J=\nu$. In practice assuming the same degree of freedom for all components of the mixture can be quite restrictive, since a single $\nu$ may not be sufficient to model the tail structure in the different components of the model. Another important point to notice is that the number of components needed to accommodate multimodality and skewness may be different from the number of components to model tail behavior. This is our motivation to propose a flexible and parsimonious mixture model that does not suffer from estimation problems and have an easier interpretation for the model components.

\subsection{The proposed model}\label{sec:2.2}

\noindent Define the $n-$dimensional response vector $\bff{Y}$, the $n\times p$ design matrix $\bff{X}$ and the $(p+1)-$dimensional coefficient vector $\bff{\beta}=(\beta_0,\beta_1,\ldots,\beta_p)^{\top}$. Consider a $J$-dimensional weight vector $\bff{w}=(w_1,\ldots, w_J)^{\top}$ and $J$ $K$-dimensional weight vectors $\bff{\dot{w}}_{j}=(\dot{w}_{j1},\ldots,\dot{w}_{jK})^{\top}$, such that $\sum_{j=1}^Jw_j=1$ and $\sum_{k=1}^K\dot{w}_{jk}=1$, $\forall\;j$; $\bff{\mu}=(\mu_1,\ldots,\mu_J)^{\top}$ $\bff{\sigma^2}=(\sigma^2_1,\ldots,\sigma^2_J)^{\top}$ and $\bff{\nu}= (\nu_1,\ldots,\nu_K)^{\top}$, with $\nu_k$ fixed and known for all $k$.

Consider the linear regression model
\begin{equation}\label{model-class}
Y_i=\beta_0 +\bff{X}_i^{\top}\bff{\beta}+\epsilon_{i}, \quad i=1,\ldots, n,
\end{equation}
where $\bff{\beta}=(\beta_1,\ldots,\beta_p)^{\top}$. We propose the following finite mixture model for the error terms distribution:
\begin{eqnarray}\label{densErro}
f(\epsilon_{i}) = \displaystyle \sum_{j=1}^J w_j\sum_{k=1}^K \dot{w}_{jk}f_{\mathcal{T}}(\mu_j, \sigma^2_j,\nu_k), \quad i=1,\ldots, n,
\end{eqnarray}
where $f_{\mathcal{T}}(\cdot)$ denotes the Student-t distribution with mean $\mu_j$, dispersion $\sigma^2_j$ and degrees of freedom $\nu_k$. Model identifiability is achieved by setting mean zero to the erros, i.e. $\ds \sum_{j=1}^J w_j\mu_j=0$.

The model in \eqref{densErro} is quite general and includes as a submodel the one propose in \citet{galimberti2014multivariate}, in a univariate context. That occurs for $J=K$ and $\bff{\dot{w}}=(\dot{w}_{1},\ldots,\dot{w}_{J})^{\top}$ being the identity matrix. Moreover, if $J=K=1$, the model in (\ref{densErro}) results in the linear regression model with Student-t errors propose by \citet{lange1989}.

Given the expressions \eqref{model-class} and \eqref{densErro}, the probability density function of the response vector $\bff{Y}$ is given by
\begin{eqnarray}\label{densY_proposto}
f(y_i) = \displaystyle \sum_{j=1}^J w_j\sum_{k=1}^K \dot{w}_{jk}f_{\mathcal{T}}(y_i|\tilde{\mu}_{ij},\sigma^2_j,\nu_k), \quad \forall i=1,\ldots, n,
\end{eqnarray}
where $\tilde{\mu}_{ij} = \beta_0+\bff{X}_i^{\top}\bff{\beta}+\mu_j$. For computational reasons, we reparametrise the model by making
$\tilde{\mu}_{ij} = \mu_j^{\ast}+\bff{X}_i^{\top}\bff{\beta}$, with the $\mu_j^{\ast}$'s the unrestricted means of the erros' mixture. The original parametrisation can be easily recovered by making $\ds\beta_0=\sum_{j=1}^J w_j\mu_j^{\ast}$ and $\mu_j = \mu_j^{\ast} - \beta_0$.

We also consider a set of auxiliary variables to easy the MCMC computations, leading to the following hierarchical representation of the proposed model:
\begin{eqnarray}
(Y_i|\bff{X}_i, Z_{ij}=1, U_i=u_i,\tilde{\mu}_{ij}, \sigma^2_j) &\stackrel{ind}{\sim}& \displaystyle \mathcal{N}\left(\tilde{\mu}_{ij},\sigma_j^2u_i^{-1}\right), \label{modeleq1}\\
   (U_i|\dot{Z}_{ijk}=1,\nu_k) &\stackrel{ind}{\sim}& \mathcal{G}\left(\frac{\nu_k}{2},\frac{\nu_k}{2}\right), \label{modeleq2} \\
   \bff{Z_i}|\bff{w} &\stackrel{iid}{\sim}& \mathcal{M}(1,w_1\ldots w_J), \label{modeleq3} \\
   \bff{\dot{Z}_{ij}}|\bff{\dot{w}}_{j} &\stackrel{ind}{\sim}& \mathcal{M}(1,\dot{w}_{j1}\ldots,\dot{w}_{jK}), \label{modeleq4}
\end{eqnarray}
where $\mathcal{N}(\cdot)$, $\mathcal{G}(\cdot)$ and $\mathcal{M}(\cdot)$ denote the Gaussian, gamma and multinomial distributions, respectively; $\bff{U}=(U_1, \ldots, U_n)^{\top}$; $\bff{Z_i}=(Z_{i1},\ldots, Z_{iJ})^{\top}$ and $\bff{\dot{Z}_{ij}}=(\dot{Z}_{ij1},\ldots,\dot{Z}_{ijK})^{\top}$ are the latent mixture component indicator variables. Variable $\bff{Z_i}$ defines the mixture component of the $i-$th individual in terms of the location and scale parameters and  $\bff{\dot{Z}_{ij}}$ defines the component in terms of the degrees of freedom parameter.

Interpretation of the model in \eqref{densY_proposto} may be facilitated by considering the following hierarchical representation of the mixture distribution in which the first level models multimodality/skewness and the second level models tail behavior.
\begin{eqnarray}
 \bff{Y}|\bff{U} &\sim&  \sum_{j=1}^J w_jf_{\mathcal{N}}(\tilde{\mu}_{ij},\sigma^2_j\bff{u}^{-1}), \label{level1}\\
  \bff{U} &\sim&  \sum_{k=1}^K \dot{w}_{k}f_{\mathcal{G}}\left(\frac{\nu_k}{2},\frac{\nu_k}{2}\right). \label{level2}
\end{eqnarray}

The mixture of gamma distributions for $U_i$ in (\ref{level2}) gives a nonparametric flavor to the model since particular choices of this distribution lead to specific marginal distributions for $\bff{Y}$ \citep[see][]{andrews1974}. The 2-level representation allows visualising the two possible sources of unobserved heterogeneity in $\bff{Y}$. The two sources of heterogeneity come, respectively, from (\ref{level1}) and (\ref{level2}).

The main advantages of the proposed model are:
\vspace{-.2cm}
\begin{enumerate}
  \item The mixture in $J$ models the multimodal and/or skewness behavior and can, therefore, have the number of mixture components fixed at reasonable values, based on an empirical analysis of the data;
  \item The separate modelling for tail bahavior may avoid over-parametrisation of the model (too many degrees of freedom parameters) due to the need for several modes to capture multimodality/skewness;
  \item Mixing $K$ different degrees of freedom, and estimating the respective weights, brings flexibility to the model in the sense that it is capable of capturing a variety of tail structures;
  \item The tail behavior is estimated without the need to estimate degrees freedom parameters. Besides avoiding the known inference difficulties associated to those parameters, our model's parsimony is less penalised with a increase in $K$.
\end{enumerate}

A particular case of the model defined in (\ref{densY_proposto}) occurs when $(\beta_0,\bff{\beta})=0$ (model without covariates) and no restriction is imposed to the $\mu_j$'s. In this case, $\bff{Y}$ has the following probability density function:
\begin{eqnarray}\label{densY2}
f(y_i) = \displaystyle \sum_{j=1}^J w_j\sum_{k=1}^K \dot{w}_{jk}f_{\mathcal{T}}(y|\mu_j,\sigma^2_j,\nu_k) \; i = 1,\ldots,n.
\end{eqnarray}
This iid mixture model can be used for cluster analysis or density estimation. Moreover, for $J=K$ and $\bff{\dot{w}}=(\dot{w}_{1},\ldots,\dot{w}_{J})^{\top}$ being the identity matrix, the model results in an ordinary mixture of Student-t distributions \citep{peel2000robust}.

\subsection{The choice of $\bff{\nu}$}\label{sec:2.3}

\begin{figure}[H]
  \centering{
  {\includegraphics[width=0.7 \textwidth]{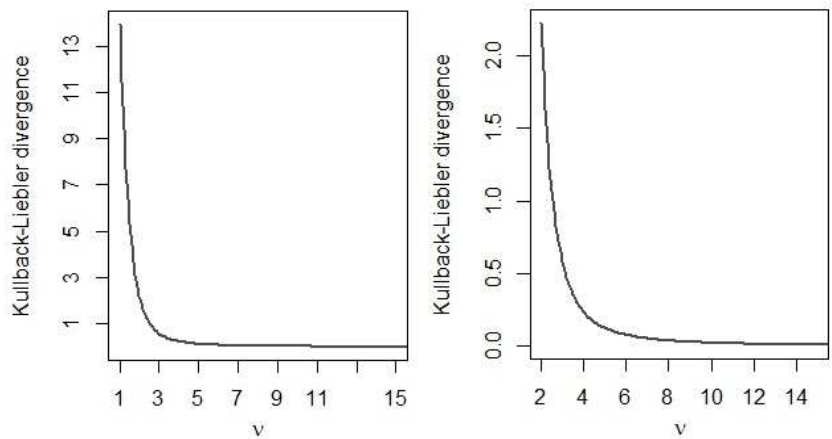}} \\
  \caption{Kullback-Liebler divergence between the standard Gaussian Student-t distributions with $\nu$ degrees of freedom.} \label{fig:1}}
\end{figure}

In this section we present a strategy on how to choose the values of the degrees of freedom parameters $\nu_k$, which are fixed in the proposed model, based on the Kullback-Liebler divergence \citep[KLD,][]{kullback1951information}. The KLD between two continuous probability measures is defined as
\begin{equation*}
    KLD (f||h) = \displaystyle \int_{-\infty}^{+\infty}f(y)\log\frac{f(y)}{h(y)}dy.
\end{equation*}
We consider the KLD between the standard Gaussian and Student-t distributions for different degrees of freedom. Figure \ref{fig:1} shows the KLD for different values of $\nu$. The similarity between those distributions when the degrees of freedom increase gives the intuition to why this is an statistically complicated parameter.

An exploratory study, not reported here, strongly suggests that a mixture of two standard Student-t distributions with d.f. $\nu_1$ and $\nu_2$ can approximate quite well a standard Student-t distribution with d.f. $\nu\in[\nu_1,\nu_2]$ if $\nu_1$ and $\nu_2$ are not too far from each other in the KLD scale. Based on that we suggest the following strategy to choose the values of $\nu_k$:
\begin{enumerate}
  \item Choose the minimum and maximum values of the $\nu_k$ parameters - $(\nu_m,\nu_M)$. The maximum should typically be between 8 and 15 and the minimum should be chosen based on how heavy the tails are believed to be. For example, one should choose this to be greater than 2 in most of the cases, so that the model has finite variance.
  \item Choose the value of $K$. Based on the exploratory study just mentioned, we suggest $K=3$ or $K=4$.
  \item Compute the values of the remaining $\nu_j$'s such that they are equally spaced in the KLD scale, i.e. $(KLD(\nu_j)-KLD(\nu_{j+1}))=(KLD(\nu_m)-KLD(\nu_M))/(K-1)$, $j=1,\ldots,K-1$. For example, for $\nu_m=2.8$, $\nu_M=14.4$ and $K=4$, we get $\bff{\nu}=(2.8,3.2,3.9,14.4)$.
\end{enumerate}


\section{Bayesian Inference}\label{sec:3}

The Bayesian model is fully specified by \eqref{modeleq1}-\eqref{modeleq4} and the prior distribution for the parameter vector $\bff{\theta}=(\bff{\mu}_j^{\ast},\bff{\sigma}^2,\bff{w}, \bff{\dot{w}}_{j}, \bff{\beta})$. Prior specification assumes independence among all the components of $\bff{\theta}$, except for $(\bff{\mu}_j^{\ast},\bff{\sigma}^2)$. The following prior distributions are adopted: $(\bff{\mu_j^{\ast},\sigma_j^{2}})\sim \mathcal{NIG}(\mu_0,\tau,\dot{\alpha},\dot{\beta})$; $\bff{w}\sim \mathcal{D}(\alpha_1,\ldots,\alpha_J)$; $\bff{\dot{w}}_{j}\sim \mathcal{D}(\dot{\alpha}_{j1},\ldots,\dot{\alpha}_{jK})$, $\forall j=1,\ldots,J$; and $\bff{\beta}\sim\mathcal{N}_p(\phi,\upsilon^2I_p)$, where $\mathcal{NIG}(\cdot)$ and $\mathcal{D}(\cdot)$ denote the normal-inverse-gamma and Dirichlet distributions, respectively.

\subsection{MCMC}\label{mcmc}

The blocking scheme for our Gibbs sampler is chosen in a way to get the larger possible blocks for which we can directly simulate from the respective full conditional distributions. The following blocks are chosen:
\begin{equation}\label{block}
 (\bff{w},\bff{\dot{w}}), \quad (\bff{U},\bff{Z},\bff{\dot{Z}}), \quad (\bff{\mu}^{\ast}, \bff{\sigma^2}), \quad \bff{\beta},
\end{equation}
where $\bff{Z}=(\bff{Z}_1,\ldots,\bff{Z}_n)^{\top}$ and $\bff{\dot{Z}}=(\bff{\dot{Z}_{1j}},\ldots,\bff{\dot{Z}_{nj}})^{\top}$ are matrices of dimensions $n\times J$ and $n\times K$, respectively.

All the full conditional distributions are derived from the joint distribution density of all random components in the model, which is given by
\begin{eqnarray*}
\displaystyle \pi(\bff{Y},\bff{U},\bff{Z},\bff{\dot{Z}},\bff{w},\bff{\dot{w}},\bff{\mu}^{\ast},\bff{\sigma^2}, \bff{\beta}, \bff{\nu})&\propto& \pi(\bff{Y}|\bff{Z}, \bff{U}, \bff{\mu}^{\ast}, \bff{\sigma^2},\bff{\beta})\pi(\bff{U}|\bff{Z}, \bff{\dot{Z}}, \bff{\nu})\pi(\bff{\dot{Z}}|\bff{Z},\bff{\dot{w}_j})\times\nonumber \\
&&\times \pi(\bff{Z}|\bff{w})\pi(\bff{w})\pi(\bff{\dot{w}})\pi(\bff{\mu}^{\ast},\bff{\sigma^2})\pi(\bff{\beta}).
\end{eqnarray*}

Details about how to sample from each of those distributions are presented in in Appendix A of the Supplementary Material.

\subsection{Prediction}\label{pred}

Prediction for unobserved configurations of the covariates is a typical aim in a regression analysis. Suppose we want to make prediction for the covariate values in the matrix $\bff{X}_{n+1}$. In an MCMC context, the output of the algorithm can be directly used to obtain a sample from the posterior predictive distribution of $Y_{n+1}$. That is achieved by adding the following steps to each iteration of the Gibbs sampler after the burn-in:\\

{\small
\begin{tabular}[!]{|l|}
\hline
\parbox[!]{14cm}{
\texttt{
\begin{itemize}
  \item Let $\left(\bff{Z}^{(m)}, \bff{\dot{Z}}^{(m)}, \bff{\sigma}^{2(m)}, \bff{\mu}^{\ast(m)}, \bff{\beta} \right)$ be the state of the chain at the $m-$th iteration. For each $m=1,2,\ldots,\tilde{M}$:
 \end{itemize}
      \begin{enumerate}
  \item Sample $\left(u_{n+1}^{(m)}|\bff{\dot{Z}}^{(m)},\bff{\nu}\right)$ from (\ref{modeleq2});
  \item Sample $Y_{n+1}^{(m)}|\bff{Z}^{(m)}\sim\mathcal{N}\left(\mu_j^{\ast(m)}+\bff{X}_{n+1}\bff{\beta}^{(m)},\sigma_j^{2(m)}\left(u_{n+1}^{(m)}\right)^{-1}\right)$.
  \end{enumerate}
}}\\  \hline
\end{tabular}} 
\section{Simulated Studies}\label{sec:4}

Two simulated studies are conducted aiming at evaluating the performance of the proposed approach. For both studies we assume $K=3$ or $K=4$ and set $\nu_m=2.8$ and $\nu_M = 14.4$, which leads to $\bff{\nu}=(2.8, 3.5, 14.4)$ and $\bff{\nu}=(2.8, 3.2, 3.9, 14.4)$.

In the first study the data is simulated from the proposed model and, in the second one, from an ordinary mixture of Skew-t distributions. We consider sample sizes $500$, $1000$ and $2500$ with objective to evaluate if exist impact of sample size when we study the tail behavior of the distribution that generated the data. Three models are fit to each simulated sample: 1. the proposed model with $K=3$ ($\texttt{Mt-p1}$); 2. the proposed model with $K=4$ ($\texttt{Mt-p2}$); 3. an ordinary mixture of Student-t distributions with 2 components in the first study and 4 in the second one ($\texttt{Mt}$). Inference for the third model is performed via MCMC by appropriately adapting our algorithm in terms of $J$, $K$ and $\bff{\dot{w}}$, and by adding a step to sample the degrees of freedom parameters. This sampling step is proposed in \citet{gonccalves2015robust}. Furthermore, the penalised complexity prior (PC prior) from \citet{simpson2017penalising} is adopted for $\bff{\nu}$ (see Appendix B of the Supplementary Material).

Comparison among the three models is performed in terms of posterior statistics of the regression coefficients and of the errors' variance, which is given by:
\begin{eqnarray*}
  V_{\epsilon}:=\mbox{Var}(\epsilon) &=& \displaystyle \sum_{j=1}^J w_j \left[ (\lambda_j- \lambda_{mix})^2 + \varsigma^2_j\right],
\end{eqnarray*}
where $\lambda_{mix}=\displaystyle \sum_{j=1}^J w_j\lambda_j$ and $\lambda_j$ and $\varsigma^2_j$ are the mean and variance of the Student-t distribution in $j-$th component of the mixture. We consider the following three statistics: bias$=E[V_{\epsilon}|\bff{y}]-V_{\epsilon}^T$, $V[V_{\epsilon}|\bff{y}]$ and $MSE=\mbox{bias}^2+V[V_{\epsilon}|\bff{y}]=E[(V_{\epsilon}-V_{\epsilon}^T)^2|\bff{y}]$, where $V_{\epsilon}^T$ is the true value of $V_{\epsilon}$.

We also define the following percentual variation measure $\bar{D}$ to compare the fitted error distribution in each model:
\begin{eqnarray*}
  \bar{D} &=& \displaystyle \frac{1}{B}\sum_{b=1}^B \Big|\frac{f^{real}(x_b)-\hat{f}(x_b)}{f^{real}(x_b)}\Big|,
\end{eqnarray*}
For a grid $\{x_b\}_{b=1}^B$ in the sample space of the error, where $f^{real}$ is to the true density of the data and $\hat{f}$ is the posterior mean of the error's density under the fitted model. The comparison is performed globally and in the tails of the distribution. In the latter, we use consider a grid of points below $1$th and above $99$th percentiles.

Initial values for $\bff{\mu^{\ast}}$, $\bff{\sigma}^2$ and $\bff{w}$ are obtained through the \texttt{R} package \texttt{mixsmsn} \citep{prates2013mixsmsn}. Matrix $\bff{\dot{w}}$ is initialised assuming the same probability for all elements and for $(\beta_0,\bff{\beta})$ we use the ordinary least square estimates. We also set the hyperparameter values $\mu_0=\bar{Y}$, $\tau= 0.005$, $\dot{\alpha}=1$ and $\dot{\beta}= 1.5$. A uniform distribution on the respective simplex is assumed for $\bff{w}$ and $\bff{\dot{w}}_{j}$.

All the MCMC chains run for $50000$ iterations with a burn-in of $10000$. A lag was defined to maximise the effective sample size of the log-posterior density. The algorithm was implemented using the \texttt{R} software \citep{teamR}.

\subsection{Errors generated from the proposed mixture model}

Data is generated from the proposed mixture model with $J=K=2$, where $\bff{w}^{\top}=(0.6, 0.4)$, $\bff{\dot{w}_j}^{\top}=(0.5, 0.5)$, $j=1,2$, $\bff{\sigma}^{2\top}=(1,0.75)$, $\bff{\nu}^{\top}=(2.8, 4)$ and $\bff{\mu}^{\top}=(-1, 1.5)$. We consider $\bff{X}_i=(1, X_{i1}, X_{i2})$, generated from $X_{i1}\sim\mathcal{N}(0,1)$ and $X_{i2}\sim\mathcal{U}(0,1)$, and $(\beta_0,\bff{\beta}^{\top})=(1,-2,1)$. Parameters $\bff{w}$ and $\bff{\mu}$ are chosen to satisfy the identifiability constraint $\displaystyle \sum_{j=1}^2w_j\mu_j=0$. The following models are fitted:
\begin{enumerate}
  \item $\texttt{Mt-p1}$: $J=2$, $K=3$ and $\bff{\nu}=(2.8, 3.5, 14.4)$;
  \item $\texttt{Mt-p2}$: $J=2$, $K=4$ and $\bff{\nu}=(2.8, 3.2, 3.9, 14.4)$;
   \item $\texttt{Mt}$: 2 mixture components.
\end{enumerate}

Table \ref{tab:2} shows that the estimates for the regression coefficients are quite similar to the true values for both models for sample size $2500$. Therefore, it is reasonable to focus on the comparison between both models on the variance of the errors. Similar results are obtained for sample sizes $500$ and $1000$.

\begin{table}[htb!]
   \caption{Posterior results for sample of size 2500 and 95\% high posterior density (HPD) when data are generated from the proposed model.}\label{tab:2}
\medskip
  \centering
\scalefont{0.7}
\begin{tabular}{l|ll|llll|l|l}
\hline
\multicolumn{1}{c|}{Model} & \multicolumn{1}{c}{$\beta_0$} & \multicolumn{1}{c|}{HPD} & \multicolumn{1}{c}{$\bff{\beta}$} & \multicolumn{1}{c|}{HPD} & \multicolumn{1}{c}{$\nu$} & \multicolumn{1}{c|}{HPD} & \multicolumn{1}{c|}{$\bff{w}$} & \multicolumn{1}{c}{$\bff{\dot{w}}$} \\
\hline
\multicolumn{1}{c|}{$\texttt{Mt-p1}$} & \multicolumn{1}{c}{1.019} & \multicolumn{1}{c|}{[0.903, 1.144]} & \multicolumn{1}{c}{-1.993} & \multicolumn{1}{c|}{[-2.060, -1.924]} & \multicolumn{2}{c|}{-} & \multicolumn{1}{c|}{ 0.612} & \multicolumn{1}{c}{(0.642, 0.244, 0.114)} \\
\multicolumn{1}{c|}{} & \multicolumn{1}{c}{} & \multicolumn{1}{c|}{} & \multicolumn{1}{c}{0.877} & \multicolumn{1}{c|}{[ 0.635, 1.114]} & \multicolumn{2}{c|}{-} & \multicolumn{1}{c|}{0.388} & \multicolumn{1}{c}{(0.497, 0.329, 0.174)} \\
\hline
\multicolumn{1}{c|}{$\texttt{Mt-p2}$} & \multicolumn{1}{c}{1.018} & \multicolumn{1}{c|}{[0.889, 1.133]} & \multicolumn{1}{c}{-1.995} & \multicolumn{1}{c|}{[-2.060, -1.926]} & \multicolumn{2}{c|}{-} & \multicolumn{1}{c|}{0.615} & \multicolumn{1}{c}{(0.465, 0.275, 0.161, 0.099)} \\
\multicolumn{1}{c|}{} & \multicolumn{1}{c}{} & \multicolumn{1}{c|}{} & \multicolumn{1}{c}{0.875} & \multicolumn{1}{c|}{[0.647, 1.134]} & \multicolumn{2}{c|}{-} & \multicolumn{1}{c|}{0.385} & \multicolumn{1}{c}{(0.339, 0.284, 0.239, 0.138)} \\
\hline
\multicolumn{1}{c|}{$\texttt{Mt}$} & \multicolumn{1}{c}{1.025} & \multicolumn{1}{c|}{[0.896, 1.140]} & \multicolumn{1}{c}{-1.995} & \multicolumn{1}{c|}{[-2.058, -1.931]} & \multicolumn{1}{c}{2.913} & \multicolumn{1}{c|}{[2.441, 3.366]} & \multicolumn{1}{c|}{0.611} & \multicolumn{1}{c}{-} \\
\multicolumn{1}{c|}{} & \multicolumn{1}{c}{} & \multicolumn{1}{c|}{} & \multicolumn{1}{c}{0.866} & \multicolumn{1}{c|}{[0.604, 1.080]} & \multicolumn{1}{c}{2.910} & \multicolumn{1}{c|}{[2.359, 3.406]} & \multicolumn{1}{c|}{0.389} & \multicolumn{1}{c}{-} \\
\hline
\end{tabular}
\end{table}

Table \ref{tab:1} shows the results concerning the comparison criteria previously described. The proposed model performs better than ordinary mixture of Student-t model for all the sample size. The posterior results for variance and MSE for \texttt{Mt} model are greatly impacted by the sample size and, even for the largest sample size, it still presents a considerable difference in the variance and MSE in comparison to \texttt{Mt-p} models. The same thing happens regarding the distances $\bar{D}$ and $\bar{D}_{tail}$. This result is explained by the fact that the posterior variance of the degrees of freedom parameters in the \texttt{Mt} model inflates the posterior variability of all the quantities that depend on those parameters. Finally, note that differences between the results for \texttt{Mt-p1} and \texttt{Mt-p2} are small enough to suggest that there is no great impact on the choice of K.

\begin{table}[htb!]
   \caption{Bias, variance and MSE of the posterior results for the error variance in each mixture model when data are generated from the proposed model}\label{tab:1}
\medskip
  \centering
\scalefont{0.7}
\begin{tabular}{c|c|ccc|cc}
\hline
\multicolumn{1}{c|}{} & \multicolumn{1}{c|}{} & \multicolumn{3}{c|}{$V_{\epsilon}$} & \multicolumn{1}{c}{} & \multicolumn{1}{c}{} \\
\multicolumn{1}{c|}{n} & \multicolumn{1}{c|}{Model} & \multicolumn{1}{c}{bias} & \multicolumn{1}{c}{var} & \multicolumn{1}{c|}{MSE } & \multicolumn{1}{c}{$\bar{D}$} & \multicolumn{1}{c}{$\bar{D}_{tail}$} \\
\hline
\multicolumn{1}{c|}{} & \multicolumn{1}{c|}{\texttt{Mt-p1}} & \multicolumn{1}{c}{-0.486} & \multicolumn{1}{c}{0.074} & \multicolumn{1}{c|}{0.313} & \multicolumn{1}{c}{0.240} & \multicolumn{1}{c}{0.285} \\
\multicolumn{1}{c|}{500} & \multicolumn{1}{c|}{\texttt{Mt-p2}} & \multicolumn{1}{c}{-0.533} & \multicolumn{1}{c}{0.065} & \multicolumn{1}{c|}{0.349} & \multicolumn{1}{c}{0.269} & \multicolumn{1}{c}{0.325} \\
\multicolumn{1}{c|}{} & \multicolumn{1}{c|}{\texttt{Mt}} & \multicolumn{1}{c}{-0.129} & \multicolumn{1}{c}{5.266} & \multicolumn{1}{c|}{5.283} & \multicolumn{1}{c}{0.325} & \multicolumn{1}{c}{0.403} \\
\hline
\multicolumn{1}{c|}{} & \multicolumn{1}{c|}{\texttt{Mt-p1}} & \multicolumn{1}{c}{-0.115} & \multicolumn{1}{c}{0.063} & \multicolumn{1}{c|}{0.076} & \multicolumn{1}{c}{0.082} & \multicolumn{1}{c}{0.095} \\
\multicolumn{1}{c|}{1000} & \multicolumn{1}{c|}{\texttt{Mt-p2}} & \multicolumn{1}{c}{-0.148} & \multicolumn{1}{c}{0.056} & \multicolumn{1}{c|}{0.078} & \multicolumn{1}{c}{0.104} & \multicolumn{1}{c}{0.130} \\
\multicolumn{1}{c|}{} & \multicolumn{1}{c|}{\texttt{Mt}} & \multicolumn{1}{c}{-0.295} & \multicolumn{1}{c}{0.338} & \multicolumn{1}{c|}{0.425} & \multicolumn{1}{c}{0.268} & \multicolumn{1}{c}{0.388} \\
\hline
\multicolumn{1}{c|}{} & \multicolumn{1}{c|}{\texttt{Mt-p1}} & \multicolumn{1}{c}{0.122} & \multicolumn{1}{c}{0.031} & \multicolumn{1}{c|}{0.045} & \multicolumn{1}{c}{0.116} & \multicolumn{1}{c}{0.155} \\
\multicolumn{1}{c|}{2500} & \multicolumn{1}{c|}{\texttt{Mt-p2}} & \multicolumn{1}{c}{0.044} & \multicolumn{1}{c}{0.032} & \multicolumn{1}{c|}{0.034} & \multicolumn{1}{c}{0.054} & \multicolumn{1}{c}{0.068} \\
\multicolumn{1}{c|}{} & \multicolumn{1}{c|}{\texttt{Mt}} & \multicolumn{1}{c}{0.404} & \multicolumn{1}{c}{0.282} & \multicolumn{1}{c|}{0.445} & \multicolumn{1}{c}{0.213} & \multicolumn{1}{c}{0.287} \\
\hline
\multicolumn{7}{l}{$\mbox{var}(\epsilon) = 3.975$} \\
\end{tabular}
\end{table}

\subsection{Errors generated from a mixture of Skew-t distributions}

Data is generated from a regression model in which the erros follow a mixture of Skew-t distributions with 2 components and $\bff{w}^{\top}=(0.6, 0.4)$, $\bff{\sigma}^{2\top}=(1,0.75)$, $\bff{\nu}^{\top}=(2.8, 4)$, $\bff{\lambda}^{\top}=(-1.5,0.8)$ (skewness parameter) and $\bff{\mu}^{\top}=(-0.8, 1.2)$. We consider $\bff{X}_i=(1, X_{i1}, X_{i2})$, generated from $X_{i1}\sim\mathcal{N}(0,1)$ and $X_{i2}\sim\mathcal{B}er(0.5)$, and $(\beta_0,\bff{\beta}^{\top})=(1,-2,1)$. The following models are fitted:
\begin{enumerate}
  \item $\texttt{Mt-p1}$: $J=4$, $K=3$ and $\nu=(2.8, 3.5, 14.4)$;
  \item $\texttt{Mt-p2}$: $J=4$, $K=4$ and $\nu=(2.8, 3.2, 3.9, 14.4)$;
  \item $\texttt{Mt}$: 4 mixture components.
\end{enumerate}

The same conclusion obtained from Tables~\ref{tab:2} and \ref{tab:1} in the first study are valid for Table~\ref{tab:4} and \ref{tab:3}, respectively.

\begin{table}[htb!]
   \caption{Posterior results for sample of size 2500 and 95\% high posterior density (HPD) when data are generated from the mixture of Skew-t.}\label{tab:4}
\medskip
  \centering
\scalefont{0.7}
\begin{tabular}{l|ll|llll|l|l}
\hline
\multicolumn{1}{c|}{Model} & \multicolumn{1}{c}{$\beta_0$} & \multicolumn{1}{c|}{HPD} & \multicolumn{1}{c}{$\bff{\beta}$} & \multicolumn{1}{c|}{HPD} & \multicolumn{1}{c}{$\nu$} & \multicolumn{1}{c|}{HPD} & \multicolumn{1}{c|}{$\bff{w}$} & \multicolumn{1}{c}{$\bff{\dot{w}}$} \\
\hline
\multicolumn{1}{c|}{} & \multicolumn{1}{c}{0.626} & \multicolumn{1}{c|}{[0.563, 0.686]} & \multicolumn{1}{c}{-2.032} & \multicolumn{1}{c|}{[-2.071, -1.987]} & \multicolumn{2}{c|}{-} & \multicolumn{1}{c|}{0.148  } & \multicolumn{1}{c}{(0.506, 0.355, 0.139)} \\
\multicolumn{1}{c|}{$\texttt{Mt-p1}$} & \multicolumn{1}{c}{} & \multicolumn{1}{c|}{} & \multicolumn{1}{c}{1.009} & \multicolumn{1}{c|}{[0.926, 1.096]} & \multicolumn{2}{c|}{-} & \multicolumn{1}{c|}{0.291} & \multicolumn{1}{c}{(0.255, 0.305, 0.440)} \\
\multicolumn{1}{c|}{} & \multicolumn{1}{c}{} & \multicolumn{1}{c|}{} & \multicolumn{1}{c}{} & \multicolumn{1}{c|}{} & \multicolumn{2}{c|}{-} & \multicolumn{1}{c|}{0.304} & \multicolumn{1}{c}{(0.237, 0.257, 0.506)} \\
\multicolumn{1}{c|}{} & \multicolumn{1}{c}{} & \multicolumn{1}{c|}{} & \multicolumn{1}{c}{} & \multicolumn{1}{c|}{} & \multicolumn{2}{c|}{-} & \multicolumn{1}{c|}{0.257} & \multicolumn{1}{c}{(0.414, 0.454, 0.132)} \\
\hline
\multicolumn{1}{c|}{} & \multicolumn{1}{c}{0.621} & \multicolumn{1}{c|}{[0.559, 0.679]} & \multicolumn{1}{c}{-2.034} & \multicolumn{1}{c|}{[-2.079, -1.992]} & \multicolumn{2}{c|}{-} & \multicolumn{1}{c|}{0.091} & \multicolumn{1}{c}{(0.309, 0.284, 0.258, 0.149)} \\
\multicolumn{1}{c|}{$\texttt{Mt-p2}$} & \multicolumn{1}{c}{} & \multicolumn{1}{c|}{} & \multicolumn{1}{c}{1.002} & \multicolumn{1}{c|}{[0.923, 1.087]} & \multicolumn{2}{c|}{-} & \multicolumn{1}{c|}{0.211} & \multicolumn{1}{c}{(0.240, 0.230, 0.253, 0.277)} \\
\multicolumn{1}{c|}{} & \multicolumn{1}{c}{} & \multicolumn{1}{c|}{} & \multicolumn{1}{c}{} & \multicolumn{1}{c|}{} & \multicolumn{2}{c|}{-} & \multicolumn{1}{c|}{0.332} & \multicolumn{1}{c}{(0.168, 0.223, 0.204, 0.405)} \\
\multicolumn{1}{c|}{} & \multicolumn{1}{c}{} & \multicolumn{1}{c|}{} & \multicolumn{1}{c}{} & \multicolumn{1}{c|}{} & \multicolumn{2}{c|}{-} & \multicolumn{1}{c|}{0.366} & \multicolumn{1}{c}{(0.388, 0.295, 0.209, 0.108)} \\
\hline
\multicolumn{1}{c|}{} & \multicolumn{1}{c}{0.633} & \multicolumn{1}{c|}{[0.572, 0.700]} & \multicolumn{1}{c}{-2.031} & \multicolumn{1}{c|}{[-2.063, -1.991]} & \multicolumn{1}{c}{7.443  } & \multicolumn{1}{c|}{[2.324, 13.485]} & \multicolumn{1}{c|}{0.174} & \multicolumn{1}{c}{-} \\
\multicolumn{1}{c|}{$\texttt{Mt}$} & \multicolumn{1}{c}{} & \multicolumn{1}{c|}{} & \multicolumn{1}{c}{1.009} & \multicolumn{1}{c|}{[0.934, 1.092]} & \multicolumn{1}{c}{7.546} & \multicolumn{1}{c|}{[2.545, 13.711]} & \multicolumn{1}{c|}{0.369} & \multicolumn{1}{c}{-} \\
\multicolumn{1}{c|}{} & \multicolumn{1}{c}{} & \multicolumn{1}{c|}{} & \multicolumn{1}{c}{} & \multicolumn{1}{c|}{} & \multicolumn{1}{c}{7.580} & \multicolumn{1}{c|}{[2.442, 13.575]} & \multicolumn{1}{c|}{0.211} & \multicolumn{1}{c}{-} \\
\multicolumn{1}{c|}{} & \multicolumn{1}{c}{} & \multicolumn{1}{c|}{} & \multicolumn{1}{c}{} & \multicolumn{1}{c|}{} & \multicolumn{1}{c}{7.534} & \multicolumn{1}{c|}{[2.168, 13.482]} & \multicolumn{1}{c|}{0.246} & \multicolumn{1}{c}{-} \\
\hline
\end{tabular}
\end{table}

We highlight the fact that the proposed methodology is efficient to simultaneously accommodate multimodality, skewness and heavy tails, with the advantage of not having to estimate the degrees of freedom parameters.

\begin{table}[htb!]
   \caption{Bias, variance and mean square error (MSE) of the posterior results for the error variance in each mixture model when data are generated from a mixture of Skew-t.}\label{tab:3}
\medskip
  \centering
\scalefont{0.7}
\begin{tabular}{l|l|lll|ll}
\hline
\multicolumn{1}{c|}{} & \multicolumn{1}{c|}{} & \multicolumn{3}{c|}{$V_{\epsilon}$} & \multicolumn{1}{c}{} & \multicolumn{1}{c}{} \\
\multicolumn{1}{c|}{n} & \multicolumn{1}{c|}{Model} & \multicolumn{1}{c}{bias} & \multicolumn{1}{c}{var} & \multicolumn{1}{c|}{MSE } & \multicolumn{1}{c}{$\bar{D}$} & \multicolumn{1}{c}{$\bar{D}_{tail}$} \\
\hline
\multicolumn{1}{c|}{} & \multicolumn{1}{c|}{\texttt{Mt-p1}} & \multicolumn{1}{c}{-0.255} & \multicolumn{1}{c}{0.134} & \multicolumn{1}{c|}{0.198} & \multicolumn{1}{c}{1.286} & \multicolumn{1}{c}{1.894} \\
\multicolumn{1}{c|}{500} & \multicolumn{1}{c|}{\texttt{Mt-p2}} & \multicolumn{1}{c}{-0.289} & \multicolumn{1}{c}{0.125} & \multicolumn{1}{c|}{0.209} & \multicolumn{1}{c}{1.353} & \multicolumn{1}{c}{2.361} \\
\multicolumn{1}{c|}{} & \multicolumn{1}{c|}{\texttt{Mt}} & \multicolumn{1}{c}{1.096} & \multicolumn{1}{c}{11.285} & \multicolumn{1}{c|}{12.486} & \multicolumn{1}{c}{1.571} & \multicolumn{1}{c}{2.634} \\
\hline
\multicolumn{1}{c|}{} & \multicolumn{1}{c|}{\texttt{Mt-p1}} & \multicolumn{1}{c}{0.103} & \multicolumn{1}{c}{0.197} & \multicolumn{1}{c|}{0.208} & \multicolumn{1}{c}{0.873} & \multicolumn{1}{c}{0.935} \\
\multicolumn{1}{c|}{1000} & \multicolumn{1}{c|}{\texttt{Mt-p2}} & \multicolumn{1}{c}{-0.061} & \multicolumn{1}{c}{0.165} & \multicolumn{1}{c|}{0.169} & \multicolumn{1}{c}{0.671} & \multicolumn{1}{c}{0.650} \\
\multicolumn{1}{c|}{} & \multicolumn{1}{c|}{\texttt{Mt}} & \multicolumn{1}{c}{0.065} & \multicolumn{1}{c}{5.158} & \multicolumn{1}{c|}{5.162} & \multicolumn{1}{c}{0.972} & \multicolumn{1}{c}{1.107} \\
\hline
\multicolumn{1}{c|}{} & \multicolumn{1}{c|}{\texttt{Mt-p1}} & \multicolumn{1}{c}{-0.036} & \multicolumn{1}{c}{0.053} & \multicolumn{1}{c|}{0.054} & \multicolumn{1}{c}{0.465} & \multicolumn{1}{c}{0.438} \\
\multicolumn{1}{c|}{2500} & \multicolumn{1}{c|}{\texttt{Mt-p2}} & \multicolumn{1}{c}{-0.052} & \multicolumn{1}{c}{0.051} & \multicolumn{1}{c|}{0.053} & \multicolumn{1}{c}{0.272} & \multicolumn{1}{c}{0.262} \\
\multicolumn{1}{c|}{} & \multicolumn{1}{c|}{\texttt{Mt}} & \multicolumn{1}{c}{-0.144} & \multicolumn{1}{c}{0.086} & \multicolumn{1}{c|}{0.107} & \multicolumn{1}{c}{0.764} & \multicolumn{1}{c}{0.918} \\
\hline
\multicolumn{7}{l}{$\mbox{var}(\epsilon)= 4.964$} \\
\end{tabular}
\end{table}

\begin{figure}[H]
 \begin{minipage}[t]{0.5\linewidth}
  \centering \includegraphics[width=0.7\textwidth]{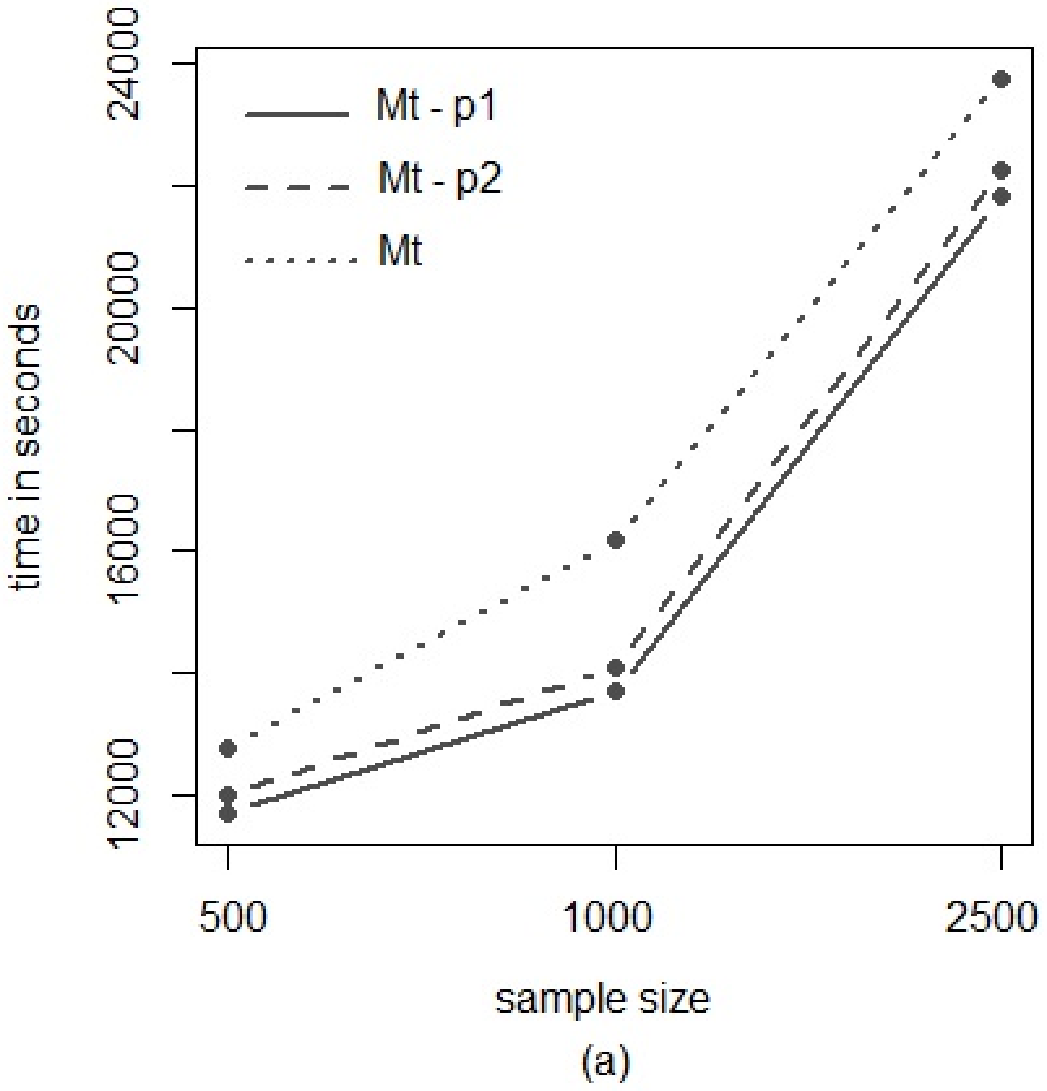}\\
 \end{minipage} \hfill
 \begin{minipage}[t]{0.5\linewidth}
  \centering \includegraphics[width=0.7\textwidth]{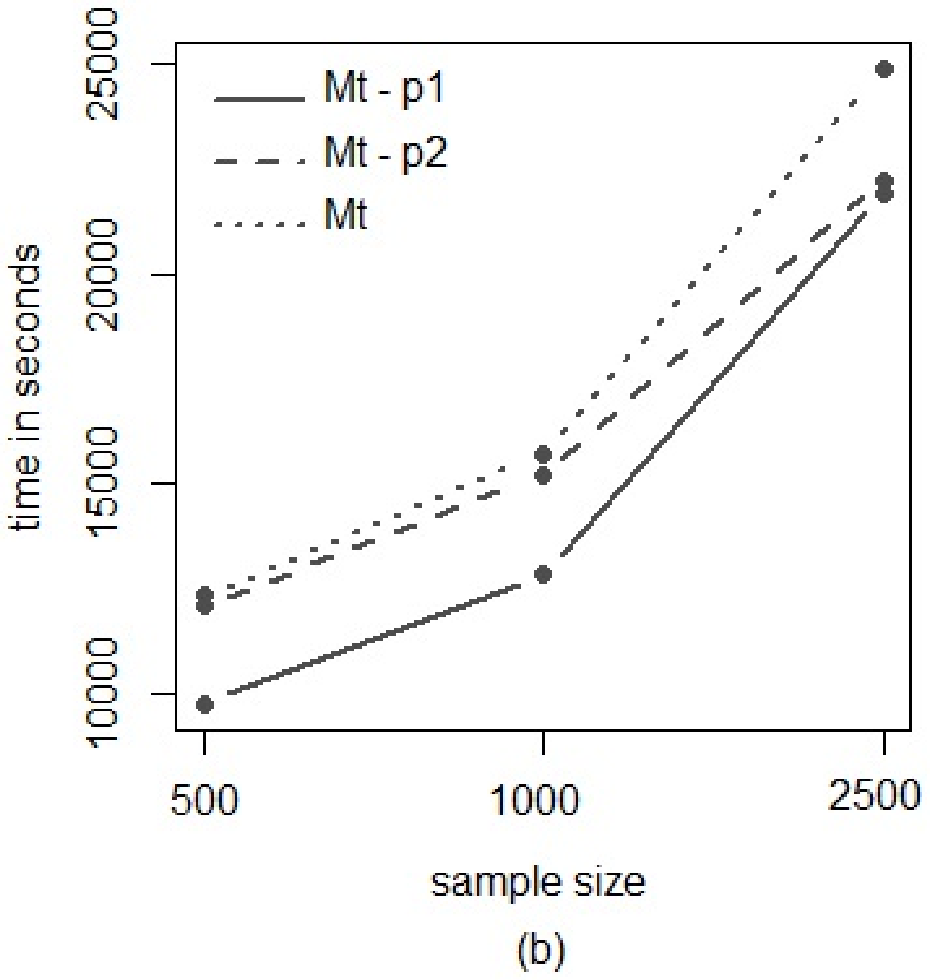}\\
 \end{minipage}
\caption{Computational cost (in seconds) of the MCMC for each model in Study 1 (a) and Study 2 (b).}\label{fig:2}
\end{figure}

Figure \ref{fig:2} shows the computational cost of the MCMC for each model as a function of the sample size, for both simulated studies. Note how the cost was uniformly smaller for the proposed model.

\section{Application}\label{sec:5}

We analyse two real data sets. The first one has been considered in several previous works and consists on the velocity of 82 galaxies (in thousands of kilometers per second) located in the \emph{Corona Borealis} constellation. The second one refers to the national health and nutrition examination (NHANES) survey conducted every year by US National Center for Health Statistics.

For the first application we assume the proposed model without covariates and, in both analysis, we also fit an ordinary mixture of Students-t distributions. We consider $K=4$ to fit the proposed model with $\bff{\nu}=(2.8, 3.2, 3.9, 14.4)$.

Model comparison is performed via DIC \citep{spiegelhalter2002bayesian}. An approximation of the DIC can be obtained using the MCMC sample $\{\bff{\theta_1,\ldots,\theta_M}\}$ from the posterior distribution. We have that $\widehat{\textrm{DIC}}= 2\overline{\textrm{D}} - \textrm{D}(\tilde{\bff{\theta}})$, where $\overline{\textrm{D}}= \displaystyle -2 \frac{1}{M} \sum_{m=1}^{M} \log f(\bff{y}|\bff{\theta}_m)$ and $\tilde{\bff{\theta}} = \textrm{E}[\bff{\theta}|\bff{y}]$.

\subsection{Galaxies velocity}

\begin{figure}[H]
   \centering\includegraphics[width=0.55\textwidth]{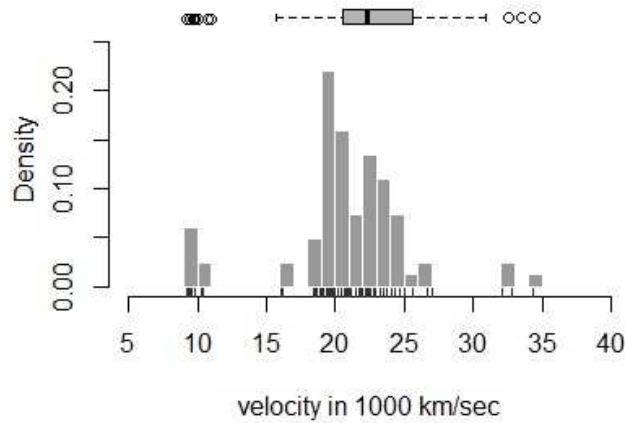}\\
\vspace{-0.2cm}
  \caption{Histogram of the galaxy data.}\label{fig:3}
\end{figure}

The data set is available in the \texttt{R} package \texttt{MASS} \citep{ripley2013package} and displayed in Figure \ref{fig:3}. This data set was previously modelled in the literature by a mixture of 6 Gaussian components \citep{carlin1995bayesian, richardson1997bayesian, stephens1997tese}. The data set has mean and variance equal to $20.83$ and exhibit some clear outliers.

Table \ref{tab:5} shows the DIC criterion for both models assuming $J=3$ and $J=4$, which refers to the total number of mixture components in the $\texttt{Mt}$ models. Based on the DIC the best fit is with $J=4$ in both models, with the \texttt{MT-p} model having the smaller DIC.

\begin{table}[htb]
  \caption{DIC selection criterion for different values of $J$ in the fitting of the \texttt{Mt-p} and \texttt{Mt} models to galaxy data.}\label{tab:5}
\scalefont{0.8}
\begin{center}
\begin{tabular}{c|c|c}
\hline
\multicolumn{1}{c|}{} & \multicolumn{2}{c}{DIC} \\
\cline{2-3}
model & $J=3$ & $J=4$ \\
\hline
\texttt{Mt-p} & $468.288$ & $\bff{427.608}$ \\
\texttt{Mt} & $471.325$ & $429.957$ \\
\hline
\end{tabular}
\end{center}
\end{table}

Table \ref{tab:6} shows some posterior statistics when $J=4$. Results are similar for both models regarding $\bff{\mu}$, $\bff{\sigma}^2$ and $\bff{w}$. The \texttt{Mt} model presented estimated values of $\nu$ between 3.88 and 4.64. For the \texttt{Mt-p} model, the values $\nu=2.8$ contributes with approximately 25\% of the weight in all four components. This result suggests that it is important to consider $\nu$ values that characterise heavy tails. Additionally, values $\nu=3.9$ and 14.4 contribute together with approximately 50\% of the weight in all components.

\begin{small}
\begin{table}[htb]
  \caption{Posterior results for the galaxy analysis. The posterior mean and the 95\% HPD credibility interval are presented.}\label{tab:6}
  \scalefont{0.5}
\begin{center}
\begin{tabular}{l|ll|ll|ll|l|l|l}
\hline
\multicolumn{1}{c|}{model} & \multicolumn{1}{c}{$\bff{\mu}$} & \multicolumn{1}{c|}{HPD} & \multicolumn{1}{c}{$\bff{\sigma}^2$} & \multicolumn{1}{c|}{HPD} & \multicolumn{1}{c}{$\bff{\nu}$} & \multicolumn{1}{c|}{HPD} & \multicolumn{1}{c|}{$\bff{w}$} & \multicolumn{1}{c|}{$\bff{\dot{w}}$} & \multicolumn{1}{c}{$\widehat{V\mbox{ar}(Y)}$} \\
\hline
\multicolumn{1}{c|}{} & \multicolumn{1}{c}{9.715} & \multicolumn{1}{c|}{[ 9.086, 10.373]} & \multicolumn{1}{c}{0.740} & [0.186, 1.648] & \multicolumn{2}{c|}{-} & \multicolumn{1}{c|}{0.086} & \multicolumn{1}{c|}{(0.233, 0.243, 0.247, 0.277)} & \multicolumn{1}{c}{} \\
\multicolumn{1}{c|}{\texttt{Mt-p}} & \multicolumn{1}{c}{19.901} & \multicolumn{1}{c|}{[19.329, 20.415]} & \multicolumn{1}{c}{0.817} & [0.220, 1.691] & \multicolumn{2}{c|}{-} & \multicolumn{1}{c|}{0.434} & \multicolumn{1}{c|}{(0.264, 0.263, 0.247, 0.226)} & \multicolumn{1}{c}{21.604} \\
\multicolumn{1}{c|}{} & \multicolumn{1}{c}{22.978} & \multicolumn{1}{c|}{[22.118, 23.682]} & \multicolumn{1}{c}{1.898} & [0.562, 3.788] & \multicolumn{2}{c|}{-} & \multicolumn{1}{c|}{0.443} & \multicolumn{1}{c|}{(0.266, 0.249, 0.255, 0.230)} & \multicolumn{1}{c}{} \\
\multicolumn{1}{c|}{} & \multicolumn{1}{c}{32.772} & \multicolumn{1}{c|}{[30.820, 34.904]} & \multicolumn{1}{c}{2.733} & [0.285, 10.144] & \multicolumn{2}{c|}{-} & \multicolumn{1}{c|}{0.037} & \multicolumn{1}{c|}{(0.246, 0.248, 0.252, 0.254)} & \multicolumn{1}{c}{} \\
\hline
\multicolumn{1}{c|}{} & \multicolumn{1}{c}{9.722} & \multicolumn{1}{c|}{[9.162, 10.462]} & \multicolumn{1}{c}{0.734} & \multicolumn{1}{c|}{[0.146, 1.709]} & \multicolumn{1}{c}{3.882} & \multicolumn{1}{c|}{[2.014, 7.008]} & \multicolumn{1}{c|}{0.086} & \multicolumn{1}{c|}{-} & \multicolumn{1}{c}{} \\
\multicolumn{1}{c|}{\texttt{Mt}} & \multicolumn{1}{c}{19.901} & \multicolumn{1}{c|}{[19.506, 20.483]} & \multicolumn{1}{c}{0.798} & \multicolumn{1}{c|}{[0.197, 1.764]} & \multicolumn{1}{c}{3.896} & \multicolumn{1}{c|}{[2.014, 7.475]} & \multicolumn{1}{c|}{0.433} & \multicolumn{1}{c|}{-} & \multicolumn{1}{c}{24.917} \\
\multicolumn{1}{c|}{} & \multicolumn{1}{c}{22.892} & \multicolumn{1}{c|}{[22.023, 23.837]} & \multicolumn{1}{c}{1.947} & \multicolumn{1}{c|}{[ 0.470, 4.041]} & \multicolumn{1}{c}{3.930} & \multicolumn{1}{c|}{[2.011, 7.565]} & \multicolumn{1}{c|}{ 0.445} & \multicolumn{1}{c|}{-} & \multicolumn{1}{c}{} \\
\multicolumn{1}{c|}{} & \multicolumn{1}{c}{32.759} & \multicolumn{1}{c|}{[29.487, 35.194]} & \multicolumn{1}{c}{2.713} & \multicolumn{1}{c|}{[0.291, 9.458]} & \multicolumn{1}{c}{4.644} & \multicolumn{1}{c|}{[2.012, 10.537]} & \multicolumn{1}{c|}{0.036} & \multicolumn{1}{c|}{-} & \multicolumn{1}{c}{} \\
\hline
\end{tabular}
\end{center}
\end{table}
\end{small}

\begin{figure}[H]
   \centering\includegraphics[width=0.8\textwidth]{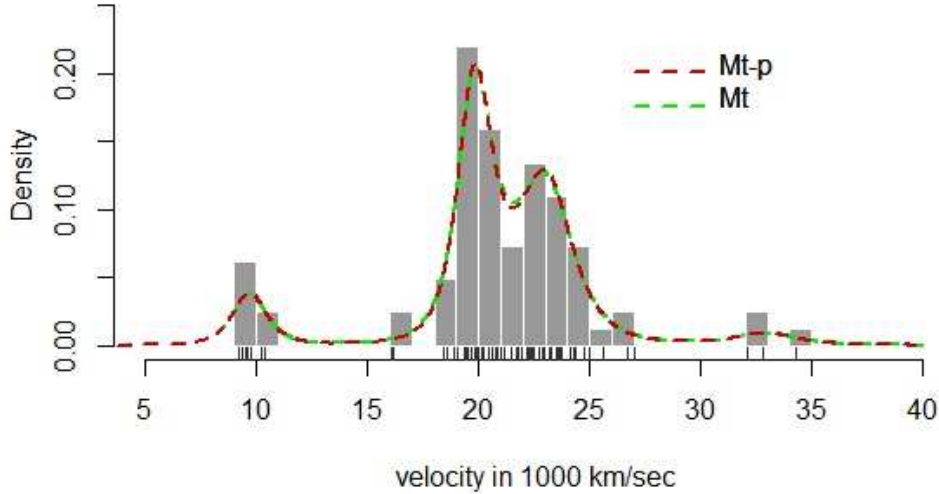}\\
  \caption{Histogram of empirical distribution of the galaxy data with fitting curves.}\label{fig:4}
\end{figure}

Figure \ref{fig:4} confirms the information in Table \ref{tab:6} that the results are quite similar for both models. Greater differences were observed for the estimates of the variance of $\bff{Y}$ and in the computational cost - $10119$ seconds for \texttt{Mt-p} and $12169$ for \texttt{Mt}). 

\subsection{US National Health and Nutrition Examination Survey}

The data set is available in the \texttt{R} package \texttt{NHANES} \citep{pruim2015nhanes} and refers to the survey carried out between 2011/2012. The data contain information on 76 variables describing demographic, physical, health, and lifestyle characteristics of 5000 participants. \citet{lin2007robust} and \citet{cabral2008bayesian} analysed the data from this study for 1999/2000 and 2001/2002 and restricted the sample to male participants only. \citet{lin2007robust} assumed a mixture of skew-t distributions to estimate the density of the participants body mass index, whereas \citet{cabral2008bayesian} used a mixture of skew-t-normal for density estimation.

\begin{figure}[H]
  \centering \includegraphics[width=0.8\textwidth]{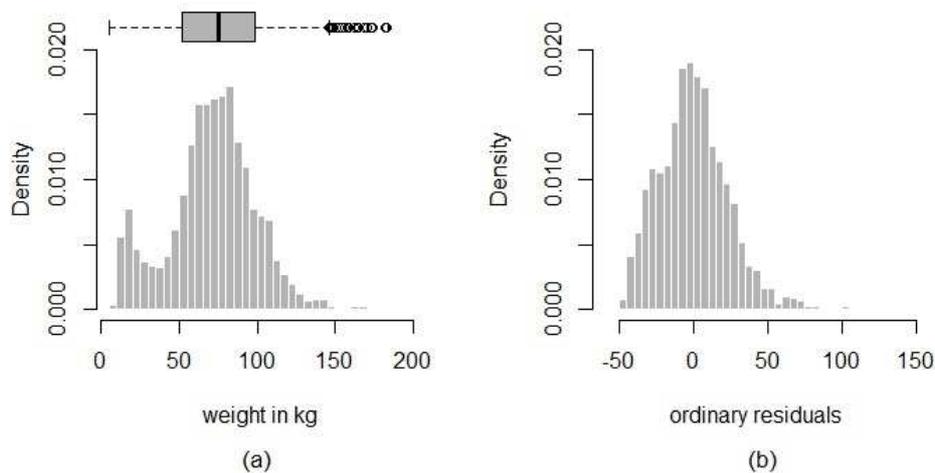}\\
\caption{(a)Histogram of empirical distribution of the weight in kilograms e (b) Histogram of ordinary residuals.}\label{fig:5}
\end{figure}

We consider the information regarding the weight in kilograms (response variable), age in years, sex and diabetes information (0-No; 1-Yes) of the participants. Participants who did not have information on at least one of the considered variables were previously removed from the database. The sample used for analysis contains 4905 participants, however 5\% of the observations are randomly selected for prediction. Thus, the final sample contains 4660 participants. Figure \ref{fig:5}(a) shows the empirical distribution of the weight of participants and exhibits a multimodal behavior. Residuals from a normal linear regression fit are positively skewed (Figure \ref{fig:5}(b)).

To model the source of unobserved heterogeneity present in Figure \ref{fig:5}(b) we applied the proposed approach to model the errors distribution and us comparer the posterior results with the mixture of t components (\texttt{Mt}). We assumed the model with $J = 2$, 3 and 4 components and based on the DIC the best fit to the data is considering a mixture with $J=4$ components (Table \ref{tab:7}).

\begin{table}[htb]
  \caption{DIC selection criterion for different values of $J$ in the fitting of the \texttt{Mt-p} and \texttt{Mt} models to NHANES data.}\label{tab:7}
\scalefont{0.8}
\begin{center}
\begin{tabular}{l|l|l|l}
\hline
\multicolumn{1}{c|}{} & \multicolumn{3}{c}{DIC} \\
\cline{2-4}
\multicolumn{1}{c|}{Model} & \multicolumn{1}{c|}{J=2} & \multicolumn{1}{c|}{J=3} & \multicolumn{1}{c}{J=4} \\
\hline
\multicolumn{1}{c|}{$\texttt{Mt-p}$} & \multicolumn{1}{c|}{$61535$} & \multicolumn{1}{c|}{$61377$} & \multicolumn{1}{c}{$\bff{56476}$} \\
\multicolumn{1}{c|}{$\texttt{Mt}$}   & \multicolumn{1}{c|}{$64822$} & \multicolumn{1}{c|}{$64403$} & \multicolumn{1}{c}{$58615$} \\
\hline
\end{tabular}
\end{center}
\end{table}

Table \ref{tab:8} shows the posterior statistics for $J=4$. Results are substantially different for both models. The posterior results suggest that the average weight of participants in the fourth component is around to 95 kg ($\mu^{\star}_4=\mu_4+\beta_0$) for the \texttt{Mt-p} model and 84 kg for the \texttt{Mt} model. The posterior mean for $\sigma^2_4$ in the \texttt{Mt} model is almost $4$ times larger than in the \texttt{Mt-p} model. Note that coefficient $\beta_3$ is less significant in model \texttt{Mt-p}. Results are also quite different for the tail behavior estimation.

\begin{small}
\begin{table}[htb]
  \caption{Posterior results for the NHANES analysis. The posterior mean and 95\% HPD credibility interval are presented.}\label{tab:8}
  \scalefont{0.7}
\begin{center}
\resizebox{\textwidth}{!}{%
\begin{tabular}{l|ll|ll|ll|ll|ll|l|l}
\hline
\multicolumn{1}{c|}{Model} & \multicolumn{1}{c}{$\bff{\mu}$} & \multicolumn{1}{c|}{HPD} & \multicolumn{1}{c}{$\bff{\sigma}^2$} & \multicolumn{1}{c|}{HPD} & \multicolumn{1}{c}{$\beta_0$} & \multicolumn{1}{c|}{HPD} & \multicolumn{1}{c}{$\bff{\beta}$} & \multicolumn{1}{c|}{HPD} & \multicolumn{1}{c}{$\nu$} & \multicolumn{1}{c|}{HPD} & \multicolumn{1}{c|}{$\bff{w}$} & \multicolumn{1}{c}{$\bff{\dot{w}}$} \\
\hline
\multicolumn{1}{c|}{} & \multicolumn{1}{c}{-30.586} & \multicolumn{1}{c|}{[-31.949, -29.303]} & \multicolumn{1}{c}{27.479 } & \multicolumn{1}{c|}{[15.473, 41.217]} & \multicolumn{1}{c}{33.154} & \multicolumn{1}{c|}{[28.241, 38.611]} & \multicolumn{1}{c}{0.642} & \multicolumn{1}{c|}{[0.609, 0.675]} & \multicolumn{2}{c|}{-} & \multicolumn{1}{c|}{0.141} & \multicolumn{1}{c}{(0.146, 0.196, 0.231, 0.427)} \\
\multicolumn{1}{c|}{$\texttt{Mt-p}$} & \multicolumn{1}{c}{-4.064} & \multicolumn{1}{c|}{[-6.263, -1.986]} & \multicolumn{1}{c}{173.528} & \multicolumn{1}{c|}{[121.409, 225.631]} & \multicolumn{1}{c}{} & \multicolumn{1}{c|}{} & \multicolumn{1}{c}{6.098} & \multicolumn{1}{c|}{[4.561, 7.742]} & \multicolumn{2}{c|}{-} & \multicolumn{1}{c|}{0.580} & \multicolumn{1}{c}{(0.027, 0.036, 0.055, 0.882)} \\
\multicolumn{1}{c|}{} & \multicolumn{1}{c}{22.012} & \multicolumn{1}{c|}{[17.254, 26.459]} & \multicolumn{1}{c}{190.682} & \multicolumn{1}{c|}{[77.107, 284.778]} & \multicolumn{1}{c}{} & \multicolumn{1}{c|}{} & \multicolumn{1}{c}{4.319} & \multicolumn{1}{c|}{[-0.305, 8.090]} & \multicolumn{2}{c|}{-} & \multicolumn{1}{c|}{0.261} & \multicolumn{1}{c}{(0.204, 0.233, 0.313, 0.250)} \\
\multicolumn{1}{c|}{} & \multicolumn{1}{c}{62.097} & \multicolumn{1}{c|}{[43.174, 70.384]} & \multicolumn{1}{c}{118.679} & \multicolumn{1}{c|}{[1.420, 393.123]} & \multicolumn{1}{c}{} & \multicolumn{1}{c|}{} & \multicolumn{1}{c}{} & \multicolumn{1}{c|}{} & \multicolumn{2}{c|}{-} & \multicolumn{1}{c|}{0.018} & \multicolumn{1}{c}{(0.245, 0.255, 0.264, 0.236)} \\
\hline
\multicolumn{1}{c|}{} & \multicolumn{1}{c}{-30.341  } & \multicolumn{1}{c|}{[-32.073, -28.768]} & \multicolumn{1}{c}{36.726} & \multicolumn{1}{c|}{[20.970, 50.130]} & \multicolumn{1}{c}{31.785} & \multicolumn{1}{c|}{[26.787, 37.580]} & \multicolumn{1}{c}{0.629} & \multicolumn{1}{c|}{[0.592, 0.664]} & \multicolumn{1}{c}{7.485} & \multicolumn{1}{c|}{[2.948, 12.535]} & \multicolumn{1}{c|}{0.162} & \multicolumn{1}{c}{-} \\
\multicolumn{1}{c|}{$\texttt{Mt}$} & \multicolumn{1}{c}{-5.202} & \multicolumn{1}{c|}{[-7.428, -3.046]} & \multicolumn{1}{c}{124.174} & \multicolumn{1}{c|}{[84.606, 169.551]} & \multicolumn{1}{c}{} & \multicolumn{1}{c|}{} & \multicolumn{1}{c}{7.146} & \multicolumn{1}{c|}{[5.260, 8.713]} & \multicolumn{1}{c}{7.456} & \multicolumn{1}{c|}{[3.056, 12.463]} & \multicolumn{1}{c|}{0.473} & \multicolumn{1}{c}{-} \\
\multicolumn{1}{c|}{} & \multicolumn{1}{c}{17.501} & \multicolumn{1}{c|}{[12.917, 22.885]} & \multicolumn{1}{c}{ 198.705} & \multicolumn{1}{c|}{[95.912, 305.835]} & \multicolumn{1}{c}{} & \multicolumn{1}{c|}{} & \multicolumn{1}{c}{4.683} & \multicolumn{1}{c|}{[0.260, 8.462]} & \multicolumn{1}{c}{7.409} & \multicolumn{1}{c|}{[2.979, 12.492]} & \multicolumn{1}{c|}{0.327} & \multicolumn{1}{c}{-} \\
\multicolumn{1}{c|}{} & \multicolumn{1}{c}{51.834} & \multicolumn{1}{c|}{[39.405, 68.744]} & \multicolumn{1}{c}{436.010} & \multicolumn{1}{c|}{[1.801, 742.636]} & \multicolumn{1}{c}{} & \multicolumn{1}{c|}{} & \multicolumn{1}{c}{} & \multicolumn{1}{c|}{} & \multicolumn{1}{c}{7.341} & \multicolumn{1}{c|}{[2.290, 12.761]} & \multicolumn{1}{c|}{0.038} & \multicolumn{1}{c}{-} \\
\hline
\end{tabular}}
\end{center}
\end{table}
\end{small}

\begin{figure}[H]
  \centering \includegraphics[width=0.95\textwidth]{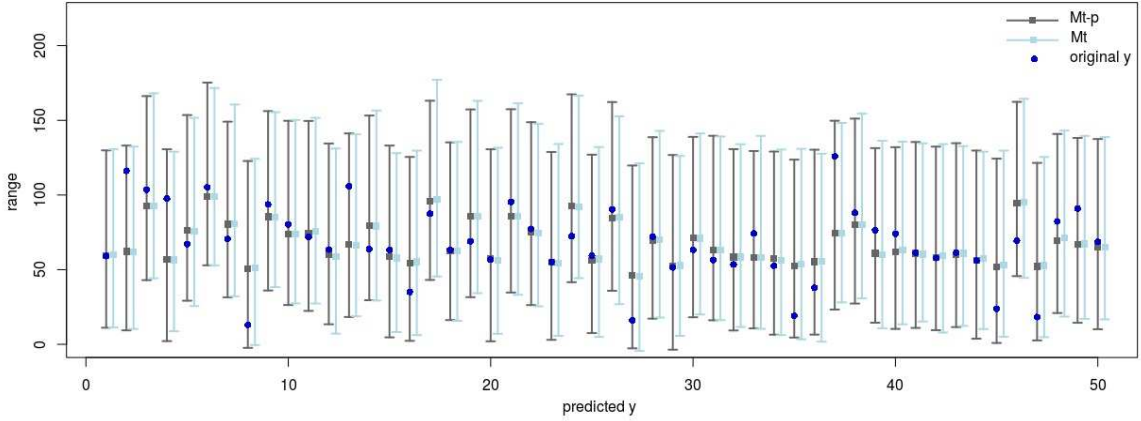}\\
\caption{HPD intervals of 99\% for the first 50 observations predicted by \texttt{Mt-p} and \texttt{Mt} models.}\label{fig:6}
\end{figure}

We also compare the models in terms of prediction capability. We consider the root of the prediction mean square error (RMSE), absolute mean error (MAE) and the relative error (RE). Results are presented in Table \ref{tab:9} and indicate a slightly better performance for the \texttt{Mt-p} model. We also consider the HPD predictive intervals which are, on average, smaller for the \texttt{Mt-p} model (see Figure \ref{fig:6} and range in Table \ref{tab:9}). The computational cost for the \texttt{Mt-p} model is $31\%$ smaller than the cost for the \texttt{Mt} model.

\begin{table}[htb]
  \caption{Prediction analysis for the \texttt{Mt-p} and \texttt{Mt} models to NHANES data.} \label{tab:9}
\scalefont{0.7}
\begin{center}
\begin{tabular}{l|llll}
\hline
\multicolumn{1}{c|}{Model} & \multicolumn{1}{c}{RMSE} & \multicolumn{1}{c}{MAE} & \multicolumn{1}{c}{RE} & \multicolumn{1}{c}{range (median)} \\
\hline
\multicolumn{1}{c|}{\texttt{Mt-p}} & \multicolumn{1}{c}{22.991} & \multicolumn{1}{c}{17.571} & \multicolumn{1}{c}{0.356} & \multicolumn{1}{c}{123.0 (122.7)} \\
\multicolumn{1}{c|}{\texttt{Mt}}   & \multicolumn{1}{c}{23.007} & \multicolumn{1}{c}{17.574} & \multicolumn{1}{c}{0.359} & \multicolumn{1}{c}{123.4 (123.1)} \\
\hline
\end{tabular}
\end{center}
\end{table}

\section{Conclusions}\label{sec:6}

This paper proposes a Bayesian model based on finite mixtures of Student-t distributions to model the errors in linear regression models. The proposed methodology considers separate structures to model multimodality/skewness and tail behavior. The two-level mixture facilitates interpretation and data modeling since it considers that the required number of components to accommodate multimodality and skewness may differ from the number of components to model the tail structure. In addition, the tail modeling does not involve degree of freedom parameters estimation, which improves the precision of estimates and the computational cost. The methodology also includes the case with no covariates for density estimation and clustering. Morevoer, the proposed MCMC algorithm may be adapted to perform inference in ordinary mixture of Student-t models including the estimation of the degrees of freedom parameters.

The performance of the proposed methodology was evaluated through simulation studies and applications to real data sets. Results illustrated the flexibility of the model to simultaneously capture the different structures presented in the errors of the regression model. It is important to emphasise that the complexity resulting from the estimation of the $K-1$ weights associated to the $\bff{\nu}$ parameters is much lower in comparison to the estimation of the degrees of freedom parameter, whose estimation process is computationally expensive and problematic.

Future work may consider estimation of the number of components $J$ and the extension for multivariate and censored data with
heavy tails. 
\section*{Appendix}
\appendix

\section{MCMC details}
The sampling step for $(\bff{U},\bff{Z},\bff{\dot{Z}})$ is based on the following factorisation:
\begin{small}
\begin{eqnarray*}\label{condU}
\pi(\bff{U},\bff{Z},\bff{\dot{Z}}|\cdot)&\propto&\displaystyle \pi(\bff{U}|\bff{Z},\bff{\dot{Z}},\bff{\nu})\pi(\bff{\dot{Z}}|\bff{Z},\bff{\dot{w}_j})\pi(\bff{Z}|\bff{w})\nonumber \\
&\propto& \displaystyle \prod_{i=1}^n \prod_{j=1}^J \left[\left(\prod_{k=1}^K \Big(\pi(U_i|\cdot)\Big)^{\dot{Z}_{ijk}} \right)^{Z_{ij}} \times
\left(\left(\prod_{k=1}^K \pi(\bff{\dot{Z}_{ij}}|\cdot) \right)^{\dot{Z}_{ijk}}\right)^{Z_{ij}} \times \Big(\pi(\bff{Z}_i|\cdot)\Big)^{Z_{ij}}\right],
\end{eqnarray*}
\end{small}
which suggests the following algorithm:
\begin{enumerate}
\item Sample the $Z_i$'s independently from
\begin{eqnarray*}
   \mathcal{M}\Big(1,\tilde{p}_{i1},\ldots,\tilde{p}_{iJ}\Big),
\end{eqnarray*}
where $\tilde{p}_{ij}=\displaystyle \frac{\tilde{r}_{ij}w_j}{\tilde{p_i}}$, with $\tilde{p}_{i}=\displaystyle \sum_{j=1}^J p_{ij}w_j$.

\item Sample the $\dot{Z}_{i Z_i}$'s independently from
\begin{eqnarray*}
  \mathcal{M}\Big(1, p_{i Z_i 1},\ldots,p_{i Z_i K} \Big),
\end{eqnarray*}
where $\displaystyle p_{ijk}=\frac{r_{ikj}\dot{w}_{jk}}{\tilde{r}_{ij}}$, with $\displaystyle r_{ikj}=\frac{\left(\sigma^2_j\right)^{-\frac{1}{2}}\left(\frac{\nu_k}{2}\right)^{\frac{\nu_k}{2}}\Gamma\left(\frac{\nu_k+1}{2}\right)}
{\Gamma\left(\frac{\nu_k}{2}\right)\left(\frac{(y_i-\tilde{\mu}_{ij})^2}{2\sigma_j^2}+\frac{\nu_k}{2}\right)^{\frac{\nu_k+1}{2}}}$ and $\displaystyle\tilde{r}_{ij}= \sum_{k=1}^K r_{ikj}\dot{w}_{j}$.

\item Sample the $U_i$'s independently from
\begin{eqnarray*}
  \mathcal{G}\left(\frac{\nu_k+1}{2},\frac{\nu_k}{2}+\frac{(y_i-\tilde{\mu}_{ij})^2}{2\sigma^2_j}\right).
\end{eqnarray*}
\end{enumerate}

The full conditional distributions of $\bff{w}$ and $\bff{\dot{w}}$ are given by
\begin{eqnarray*}
(\bff{w}|\cdot)&\sim& \displaystyle \mathcal{D}ir\left(\alpha_1+n_1,\ldots,\alpha_J+n_J \right),
\end{eqnarray*}

\begin{eqnarray*}
(\bff{\dot{w}_j}|\cdot)&\sim& \displaystyle \mathcal{D}ir\left(\dot{\alpha}_{j1}+n_{j1},\ldots,\dot{\alpha}_{jK}+n_{jK} \right), \quad \forall j=1,\ldots,J.
\end{eqnarray*}

\hspace{\parindent}The full conditional distributions of $(\bff{\mu}_j^{\ast},\bff{\sigma}^2)$ and $\bff{\beta}$ are given by:
\begin{eqnarray*}\label{condsigma}
(\bff{\sigma^2}|\cdot) &\sim& \displaystyle \mathcal{GI}\left(\alpha^{\ast}, \dot{\beta}^{\ast}\right),\\ \label{condmu}
(\bff{\mu}|\cdot)&\sim& \displaystyle \mathcal{N}\left(\mu_0^{\ast}, \frac{\bff{\sigma^2}}{\tau^{\ast}}\right),
\end{eqnarray*}
\noindent where $(\bff{\mu}_j^{\ast},\bff{\sigma}^2)\sim \mathcal{NIG}\left(\mu_0^{\ast}, \tau^{\ast}, \alpha^{\ast}, \dot{\beta}^{\ast}\right)$ with \\
$\mu_0^{\ast} = \displaystyle \frac{\Big(\sum_{i=1}^{n_1}U_{i}Y_{i}+\ldots+\sum_{i=1}^{n_J}U_{i}Y_{i}\Big) +\tau\mu_0}{\Big(\sum_{i=1}^{n_1}U_{i}+\ldots+\sum_{i=1}^{n_J}U_{i}\Big) + \tau}$; \\
\begin{flushleft}
$\tau^{\ast} = \displaystyle \Big(\sum_{i=1}^{n_1}U_{i} + \ldots +\sum_{i=1}^{n_J}U_{i}\Big) + \tau \quad; \quad \alpha^{\ast} = \displaystyle \dot{\alpha} + \frac{\sum_{i=1}^J n_1+\ldots+n_j}{2}; \quad$ and
\end{flushleft}

\begin{footnotesize}
\begin{flushleft}
$\dot{\beta}^{\ast} = \frac{1}{2}\left[\Bigg(\Big(\sum_{i=1}^{n_1}U_{i}Y_{i}^2+ \ldots+ \sum_{i=1}^{n_J}U_{i}Y_{i}^2\Big) + 2\dot{\beta}+ \tau\mu_0^2\Bigg)  -\frac{\Bigg(\Big(\sum_{i=1}^{n_1}U_{i}Y_{i}+\ldots+\sum_{i=1}^{n_J}U_{i}Y_{i}\Big)+\tau\mu_0\Bigg)^2}{\Big(\sum_{i=1}^{n_1}U_{i}+\ldots+\sum_{i=1}^{n_J}U_{i}\Big) + \tau}\right].$
\end{flushleft}
\end{footnotesize}

\begin{eqnarray*}
  (\bff{\beta}|\cdot) &\sim& \displaystyle \mathcal{N}_p\Bigg(\displaystyle \bff{\Sigma}_{\beta}\Big[(\upsilon^2I_p)^{-1}\theta + \frac{(\sqrt{ \bff{u}}\odot\bff{X}_i)^{\top}(\sqrt{\bff{u}}\odot\bff{Y}_i)}{\bff{\sigma}^2}\Big], \bff{\Sigma}_{\beta} \Bigg),
\end{eqnarray*}
\noindent where $\bff{\Sigma}_{\beta} = \displaystyle \Big[(\upsilon^2I_p)^{-1} + \frac{(\sqrt{\bff{u}}\odot\bff{X}_i)^{\top}(\sqrt{\bff{u}}\odot\bff{X}_i)}{\bff{\sigma}^2}\Big]^{-1}$; $\sqrt{\bff{u}}$ is the $n-$dimensional
vector with entries $\sqrt{u_i}$; $\odot$ is the Hadamard product and $I_p$ is the identity matrix with dimension $p$.

\vspace{0.3cm}

\section{PC priors for $\bff{\nu}$}\label{pcPrior}

The PC prior from \citet{simpson2017penalising} was constructed to prefer a simpler model $h$ and penalise the more complex one $f$. To do so, it defines a measure of complexity $d(f||h)(\bff{\nu})= d(\bff{\nu}) = \sqrt{2KLD(f||h)}$, based on the Kullback-Leibler divergence. The exponential prior is set for the measure $d(\bff{\nu})$ and thus the prior distribution of $\bff{\nu}$ is given by
$$\pi(\bff{\nu}) = \lambda \exp(-\lambda d(\bff{\nu})) \left|\frac{\partial d(\bff{\nu})}{\partial \bff{\nu}}\right|.$$
A nice feature of this prior is that the selection of an appropriate $\lambda$ is done by allowing the researcher to control the prior tail behavior of the model. For more details see \citet{simpson2017penalising}.

\bibliographystyle{elsarticle-harv}
\bibliography{reference}  

\begin{thebibliography}{25}
\expandafter\ifx\csname natexlab\endcsname\relax\def\natexlab#1{#1}\fi
\expandafter\ifx\csname url\endcsname\relax
  \def\url#1{\texttt{#1}}\fi
\expandafter\ifx\csname urlprefix\endcsname\relax\def\urlprefix{URL }\fi

\bibitem[{Andrews and Mallows(1974)}]{andrews1974}
Andrews, D.~F., Mallows, C.~L., 1974. Scale mixtures of normal distributions.
  Journal of the Royal Statistical Society. Series B - Methodological
  \textbf{36}~(1), 99--102.

\bibitem[{Bartolucci and Scaccia(2005)}]{bartolucci2005}
Bartolucci, F., Scaccia, L., 2005. The use of mixtures for dealing with
  non-normal regression errors. Computational Statistics \& Data Analysis
  \textbf{48}~(4), 821--834.

\bibitem[{Benites et~al.(2016)Benites, Maehara, and Lachos}]{benites2016}
Benites, L., Maehara, R., Lachos, V.~H., 2016. Linear regression models with
  mixture of skew heavy-tailed errors. Preprint 06/2016, Universidade Estadual
  de Campinas.

\bibitem[{Cabral et~al.(2008)Cabral, Bolfarine, and
  Pereira}]{cabral2008bayesian}
Cabral, C.~B., Bolfarine, H., Pereira, J. R.~G., 2008. Bayesian density
  estimation using skew student-t-normal mixtures. Computational Statistics \&
  Data Analysis \textbf{52}~(12), 5075--5090.

\bibitem[{Carlin and Chib(1995)}]{carlin1995bayesian}
Carlin, B.~P., Chib, S., 1995. Bayesian model choice via markov chain monte
  carlo methods. \emph{Journal of the Royal Statistical Society, series B
  (Methodological)} 57, 473--484.

\bibitem[{Fernandez and Steel(1999)}]{1999fernandez}
Fernandez, C., Steel, M. F.~J., 1999. Multivariate student-t regression models:
  Pitfalls and inference. Biometrika \textbf{86}~(1), 153--167.

\bibitem[{Fonseca et~al.(2008)Fonseca, Ferreira, and
  Migon}]{fonseca2008objective}
Fonseca, T.~C., Ferreira, M. A.~R., Migon, H.~S., 2008. Objective bayesian
  analysis for the student-t regression model. Biometrika \textbf{95}~(2),
  325--333.

\bibitem[{Fr{\"u}hwirth-Schnatter(2006)}]{sylvia2006finite}
Fr{\"u}hwirth-Schnatter, S., 2006. Finite mixture and Markov switching models:
  Modeling and applications to random processes. Springer Science \& Business
  Media.

\bibitem[{Galimberti and Soffritti(2014)}]{galimberti2014multivariate}
Galimberti, G., Soffritti, G., 2014. A multivariate linear regression analysis
  using finite mixtures of t distributions. Computational Statistics \& Data
  Analysis \textbf{71}, 138--150.

\bibitem[{Gon{\c{c}}alves et~al.(2015)Gon{\c{c}}alves, Prates, and
  Lachos}]{gonccalves2015robust}
Gon{\c{c}}alves, F.~B., Prates, M.~O., Lachos, V.~H., 2015. Robust bayesian
  model selection for heavy-tailed linear regression using finite mixtures.
  \emph{arXiv preprint arXiv:1509.00331}.

\bibitem[{Kullback and Leibler(1951)}]{kullback1951information}
Kullback, S., Leibler, R.~A., 1951. On information and sufficiency. The annals
  of mathematical statistics \textbf{22}~(1), 79--86.

\bibitem[{Lange et~al.(1989)Lange, Little, and Taylor}]{lange1989}
Lange, K.~L., Little, R.~J., Taylor, J.~M., 1989. Robust statistical modeling
  using the t distribution. Journal of the American Statistical Association
  \textbf{84}~(408), 881--896.

\bibitem[{Lin et~al.(2007)Lin, Lee, and Hsieh}]{lin2007robust}
Lin, T., Lee, J., Hsieh, W., 2007. Robust mixture modeling using the skew-t
  distribution. Statistics and Computing \textbf{17}~(2), 81--92.

\bibitem[{McLachlan and Peel(2000)}]{mclachlan2000finite}
McLachlan, G.~J., Peel, D., 2000. Finite mixture models. John Wiley \& Sons.

\bibitem[{Peel and McLachlan(2000)}]{peel2000robust}
Peel, D., McLachlan, G.~J., 2000. Robust mixture modelling using the t
  distribution. Statistics and computing \textbf{10}~(4), 339--348.

\bibitem[{Prates et~al.(2013)Prates, Lachos, and Cabral}]{prates2013mixsmsn}
Prates, M.~O., Lachos, V.~H., Cabral, C., 2013. mixsmsn: Fitting finite mixture
  of scale mixture of skew-normal distributions. Journal of Statistical
  Software \textbf{54}~(12), 1--20.

\bibitem[{Pruim(2015)}]{pruim2015nhanes}
Pruim, R., 2015. Nhanes: Data from the us national health and nutrition
  examination study.

\bibitem[{{R Core Team}(2017)}]{teamR}
{R Core Team}, 2017. R: A Language and Environment for Statistical Computing. R
  Foundation for Statistical Computing, Vienna, Austria.
\newline\urlprefix\url{https://www.R-project.org/}

\bibitem[{Richardson and Green(1997)}]{richardson1997bayesian}
Richardson, S., Green, P., 1997. On bayesian analysis of mixtures with an
  unknown number of components. Journal of the Royal Statistical Society.
  Series B - Methodological \textbf{59}~(4), 731--792.

\bibitem[{Ripley et~al.(2013)Ripley, Venables, Bates, Hornik, Gebhardt, Firth,
  and Ripley}]{ripley2013package}
Ripley, B., Venables, B., Bates, D.~M., Hornik, K., Gebhardt, A., Firth, D.,
  Ripley, M.~B., 2013. Package mass.

\bibitem[{Simpson et~al.(2017)Simpson, Rue, Riebler, Martins, S{\o}rbye,
  et~al.}]{simpson2017penalising}
Simpson, D., Rue, H., Riebler, A., Martins, T.~G., S{\o}rbye, S.~H., et~al.,
  2017. Penalising model component complexity: A principled, practical approach
  to constructing priors. Statistical Science 32~(1), 1--28.

\bibitem[{Soffritti and Galimberti(2011)}]{soffritti2011multivariate}
Soffritti, G., Galimberti, G., 2011. Multivariate linear regression with
  non-normal errors: A solution based on mixture models. Statistics and
  Computing \textbf{21}~(4), 523--536.

\bibitem[{Spiegelhalter et~al.(2002)Spiegelhalter, Best, Carlin, and Van
  Der~Linde}]{spiegelhalter2002bayesian}
Spiegelhalter, D.~J., Best, N.~G., Carlin, B.~P., Van Der~Linde, A., 2002.
  Bayesian measures of model complexity and fit. Journal of the Royal
  Statistical Society: Series B (Statistical Methodology) \textbf{64}~(4),
  583--639.

\bibitem[{Stephens(1997)}]{stephens1997tese}
Stephens, M., 1997. Bayesian methods for mixtures of normal distributions.
  Ph.D. thesis, University of Oxford.

\bibitem[{Villa et~al.(2014)Villa, Walker, et~al.}]{villa2014objective}
Villa, C., Walker, S.~G., et~al., 2014. Objective prior for the number of
  degrees of freedom of at distribution. Bayesian Analysis \textbf{9}~(1),
  197--220.

\end{thebibliography}

\end{document}